\documentclass[a4paper]{article}
\usepackage{CJK}
\usepackage[paperwidth=185mm,paperheight=230mm,textheight=190mm,textwidth=145mm,left=20mm,right=20mm, top=25mm, bottom=20mm]{geometry}
\usepackage[CJKbookmarks, colorlinks, bookmarksnumbered=true,pdfstartview=FitH,linkcolor=blue,citecolor=green]{hyperref}
\usepackage{amsmath,amssymb}
\usepackage{amsthm}
\usepackage{calc}
\usepackage{graphicx}
\usepackage{supertabular}
\usepackage{longtable}
\usepackage{float}
\usepackage{color}
\usepackage{enumerate}
\usepackage{colortbl,booktabs}
\usepackage{multirow}
\newtheorem{theorem}{Theorem}

\newtheorem{assumption}{Assumption}

\usepackage{natbib}
\usepackage{url}
\usepackage{bm}
\usepackage{enumitem}
\usepackage{authblk}

\def\wh{\widehat}
\def\wt{\widetilde}

\begin{document}

\title{ Locally sparse estimator of generalized varying coefficient model for asynchronous longitudinal data }
\author[a]{{\fontsize{12pt}{18pt}\selectfont Rou Zhong}}
\author[b]{{\fontsize{12pt}{0.5em}\selectfont Chunming Zhang}}
\author[a]{{\fontsize{12pt}{0.5em}\selectfont Jingxiao Zhang} \thanks{zhjxiaoruc@163.com}}
\affil[a]{{\emph\fontsize{12pt}{0.5em}\selectfont Center for Applied Statistics, School of Statistics, Renmin University of China}}
\affil[b]{{\emph\fontsize{12pt}{0.5em}\selectfont Department of Statistics, University of Wisconsin-Madison}}
\date{}
\maketitle

\begin{abstract}

In longitudinal study, it is common that response and covariate are not measured at the same time, which complicates the analysis to a large extent. In this paper, we take into account the estimation of generalized varying coefficient model with such asynchronous observations. A penalized kernel-weighted estimating equation is constructed through kernel technique in the framework of functional data analysis. Moreover, local sparsity is also considered in the estimating equation to improve the interpretability of the estimate. We extend the iteratively reweighted least squares (IRLS) algorithm in our computation. The theoretical properties are established in terms of both consistency and sparsistency, and the simulation studies further verify the satisfying performance of our method when compared with existing approaches. The method is applied to an AIDS study to reveal its practical merits.

\end{abstract}

\section{ Introduction }

Generalized varying coefficient model \citep{hastie1993varying, cai2000efficient} allows the coefficients to vary over time, which broadens the application of regression models to a large extent. Specifically, the model can be expressed as
\begin{align}
E \{ Y(t) | X(t) \} = g \{ \beta_0(t) + \beta_1(t) X(t) \}, t \in \mathcal{T}, \label{model}
\end{align}
where $Y(t)$ is the response, $X(t)$ is the covariate, $g ( \cdot )$ is a known, strictly increasing and continuously twice-differentiable link function, $\beta_0(t)$ is the intercept function, $\beta_1(t)$ is the varying coefficient function and $\mathcal{T}$ is a bounded and closed interval.
In this paper, we tend to develop a new estimating method for generalized varying coefficient model with longitudinal measurements from the perspective of functional data.

In practice, for longitudinal observations, it is often happened that covariate and response are not measured at the same time within each subject. Such asynchronous observations make the analysis more complicated and there are mainly two types of approaches to solve such issue in the existing work. The first type is two-step and based on synchronizing measurements of covariate and response. For example, \citet{xiong2010binning} proposed a binning method to align the measurement times so that traditional longitudinal modeling can be used. Moreover, functional principal component analysis (FPCA) is employed in \citep{csenturk2013modeling} to synchronize the data. This type of methods is not ideal enough, since the actual data used for modeling is obtained from estimation and errors from each step will accumulate. The second type is implemented through imposing kernel weight according to the observation time difference between covariate and response. This kind of methods is more appealing, since it makes full use of the whole data. \citet{cao2015regression} constructed a kernel weighted estimating equation for generalized linear model and generalized varying coefficient model. \citet{cao2016last} developed a weighted last observation carried forward (LOCF) method. Furthermore, \citet{chen2017analysis} applied the kernel weighting technique to partially linear models. \citet{li2020regression} considered models with longitudinal functional covariate. \citet{sun2021regression} investigated the cases where the observation times are informative. Most of the above kernel methods only adapt to model with time invariant coefficients, and only \citet{cao2015regression} thought about the approach of generalized varying coefficient model. However, the varying coefficients are estimated point by point, which can be time-consuming and lacks integrity. Therefore, a new estimating method is quite essential.

Interpretation of the varying coefficient function $\beta_1(t)$ is a vital part of the analysis. Moreover, interpretability can be improved through introducing local sparsity, which means the curve can be strictly equal to zero in some subintervals. In the existing work, local sparsity can be achieved by imposing sparseness penalty and was studied for different models. For example, \citet{james2009functional}, \citet{zhou2013functional} and \citet{lin2017locally} developed locally sparse estimator for scalar-on-function regression model. \citet{tu2020estimation} employed a group bridge approach to obtain locally sparse estimates for varying coefficient model. \citet{fang2020smooth} generalized the method in \citep{lin2017locally} to the cases where the response is multivariate. Function-on-function regression model and function-on-scalar regression model were also taken into account by \citet{centofanti2020smooth} and \citet{wang2020functional} respectively.
However, to the best of our knowledge, local sparsity has not been considered for generalized varying coefficient model.

In this paper, approaches in functional data analysis (FDA) is utilized, since longitudinal data can be seen as functional data in sparse design and FDA is more effective than pointwise methods. The goal is to propose a novel method that can be applied to the asynchronous data and can produce more interpretable estimates. Specifically, we construct a new kernel-weighted estimating equation with penalty on both roughness and sparseness. To solve the estimating equation, we extend the iteratively reweighted least squares (IRLS) method to our issue and design an innovative algorithm for the computation. The selection of tuning parameters is also considered. We generalize the extended Bayesian information criterion (EBIC) in \citep{chen2008extended, chen2012extended}, to make it adapt to the asynchronous data so that the roughness parameter and sparseness parameter can be chosen accordingly. Moreover, the number of basis functions is selected through cross-validation (CV). The proposed method for generalized varying coefficient model is called LocKer, as we can get locally sparse estimator of $\beta_1(t)$ from it and the kernel technique is used in the procedure. Furthermore, theoretical properties are also explored in this paper.

The contributions of our work are three-fold. First, we study generalized varying coefficient model in the framework of FDA with the consideration of both asynchronous issue and local sparsity. The problem is of high practical relevance, since it can facilitate the improvement of accuracy, utility and interpretability. Second, the newly presented algorithm can be implemented through the \textsf{R} package \texttt{LocKer} that we developed, and the package is already available on \url{https://CRAN.R-project.org/package=LocKer}. Third, we provide asymptotic properties of our method in two aspects, including consistency and sparsistency.

The paper is set out as follows. In Section \ref{SecMethod}, we illustrate the construction of the penalized kernel-weighted estimating equation and develop a computation algorithm for the proposed LocKer method. Consistency and sparsistency of the estimator are discussed in Section \ref{SecTheory}. Simulation studies are conducted in Section \ref{SecSim}, in which we explore both accuracy and zero-valued subintervals identifying ability of the method. We apply our method to an AIDS study in Section \ref{SecReal}. In Section \ref{SecDiscussion}, we conclude this paper with discussion and list some possible extension.

\section{Methodology}\label{SecMethod}

\subsection{Estimating equation}\label{SecEE}

Suppose that there are $n$ independent subjects in the study. For the $i$-th subject, let $Y_i(t)$ and $X_i(t)$ be the realization of response process $Y(t)$ and covariate process $X(t)$, respectively. However, only longitudinal measurements are obtained. In specific, for $i = 1, \ldots, n$, we observe
\begin{align}
Y_i(T_{ij}), j = 1, \ldots, L_i, \quad X_i(S_{ik}), k = 1, \ldots, M_i, \nonumber
\end{align}
where $T_{ij}$ is the $j$-th observation time of the response, $S_{ik}$ is the $k$-th observation time of the covariate, $L_i$ is the observation size of the response and $M_i$ is the observation size of the covariate. Refer to \citet{cao2015regression}, the observation times can be seen as generating from a bivariate counting process
\begin{align}
N_i(t, s) = \sum_{j = 1}^{L_i} \sum_{k = 1}^{M_i} I(T_{ij} \leq t, S_{ik} \leq s), \nonumber
\end{align}
where $I(\cdot)$ is the indicator function.

To estimate $\beta_0(t)$ and $\beta_1(t)$ in (\ref{model}), we employ the following basis approximation
\begin{align}
\beta_0(t) = \sum_{l = 1}^L B_l(t) \gamma_l^{(0)} = \textbf{B}(t)^\top \bm{\gamma}^{{(0)}}, \quad \beta_1(t) = \sum_{l = 1}^L B_l(t) \gamma_l^{(1)} = \textbf{B}(t)^\top \bm{\gamma}^{{(1)}}, \nonumber
\end{align}
where $\{B_l(t), l = 1, \ldots, L\}$ are the B-spline basis functions with degree $d$ and $K$ interior knots, $\gamma_l^{(0)}$ and $\gamma_l^{(1)}$ are the corresponding coefficients of $\beta_0(t)$ and $\beta_1(t)$, $\textbf{B}(t) = (B_1(t), \ldots, B_L(t))^\top$, $\bm{\gamma}^{(0)} = (\gamma_1^{(0)}, \ldots, \gamma_L^{(0)})^\top$, $\bm{\gamma}^{(1)} = (\gamma_1^{(1)}, \ldots, \gamma_L^{(1)})^\top$, and $L = K + d + 1$ is the number of basis functions. Here B-spline basis functions are applied, and \citet{zhong2021sparse} explained the reasons for the wide use of B-spline basis in local sparse estimation. Let $\bm{\gamma} = (\bm{\gamma}^{{(0)}\top}, \bm{\gamma}^{{(1)}\top})^\top$, $\wt{X}_l(t) = X(t) B_l(t)$ and $\wt{\textbf{X}}(t) = (\wt{X}_1(t), \ldots, \wt{X}_L(t))^\top$. Then the generalized varying coefficient model (\ref{model}) can be approximated by
\begin{align}
E \{ Y(t) | X(t) \} = g \Big \{ \sum_{l = 1}^L B_l(t) \gamma_l^{(0)} + \sum_{l = 1}^L \wt{X}_l(t) \gamma_l^{(1)} \Big \} = g \Big \{ \wt{\textbf{X}}^{\star}(t)^\top \bm{\gamma} \Big \}, \nonumber
\end{align}
where $\wt{\textbf{X}}^{\star}(t) = (\textbf{B}(t)^\top, \wt{\textbf{X}}(t)^\top)^\top$. Therefore, we can get the estimates of $\beta_0(t)$ and $\beta_1(t)$ through the estimation of $\bm{\gamma}$. To this end, we construct the following penalized kernel-weighed estimating equation
\begin{align}
U_n(\bm{\gamma}) = &\frac{1}{N_0} \sum_{i = 1}^n \sum_{j = 1}^{L_i} \sum_{k = 1}^{M_i} K_h(T_{ij} - S_{ik}) \wt{\textbf{X}}_i^{\star}(S_{ik}) \Big [Y_i({T_{ij}}) - g \Big \{ \wt{\textbf{X}}_i^{\star}(S_{ik})^\top \bm{\gamma} \Big \} \Big ] \nonumber \\
&- \textbf{V}_{\rho_0, \rho_1} \bm{\gamma} - \frac{\partial \mbox{PEN}_{\lambda}(\bm{\gamma})}{\partial \bm{\gamma}} = \textbf{0}, \label{EE_raw}
\end{align}
where $N_0 = \sum_{i = 1}^{n} L_iM_i$, $\textbf{V}_{\rho_0, \rho_1} = \mbox{diag}(\rho_0 \textbf{V}, \rho_1 \textbf{V})$, $\textbf{V} = \int_{\mathcal{T}} \textbf{B}^{(2)}(t) \textbf{B}(t)^{(2)\top} dt$, $\textbf{B}^{(2)}(t)$ is the second derivative of $\textbf{B}(t)$, $\rho_0$ and $\rho_1$ are the roughness parameters for $\beta_0(t)$ and $\beta_1(t)$, $K_h(t) = K(t/h)/h$, $K(t)$ is a symmetric kernel function, $h$ is the bandwidth, $\mbox{PEN}_{\lambda}(\bm{\gamma})$ is the sparseness penalty for $\beta_1(t)$, $\lambda$ is the sparseness parameter and $\textbf{0}$ is a zero-valued vector with length $2L$. Here we use $h = \max (\tau_{0.95}, 0.01)$ as the bandwidth, where $\tau_{0.95}$ is the 0.95-quantile of $\min_{j, k}|T_{ij} - S_{ik}|$. For the first term in (\ref{EE_raw}), we consider all possible pairs of response and covariate measurements, with the kernel weights to control the effect of various pairs, such that measurements with close observation times can be emphasized. The second term is the derivative of the roughness penalty, which is defined as
\begin{align}
&\frac{\rho_0}{2} \int_{\mathcal{T}} \{ \beta_0^{(2)} (t) \}^2 dt + \frac{\rho_1}{2} \int_{\mathcal{T}} \{ \beta_1^{(2)} (t) \}^2 dt \nonumber \\
=& \frac{\rho_0}{2} \bm{\gamma}^{{(0)}\top} \textbf{V} \bm{\gamma}^{{(0)}} + \frac{\rho_1}{2} \bm{\gamma}^{{(1)}\top} \textbf{V} \bm{\gamma}^{{(1)}} = \frac{1}{2} \bm{\gamma}^\top \textbf{V}_{\rho_0, \rho_1} \bm{\gamma}, \nonumber
\end{align}
where $\beta_0^{(2)} (t)$ and $\beta_1^{(2)} (t)$ are the second derivatives of $\beta_0(t)$ and $\beta_1(t)$. The third term is the derivative of the sparseness penalty $\mbox{PEN}_{\lambda}(\bm{\gamma})$, the expression of which is provided in Section \ref{SecSpar}. Through the computation of (\ref{EE_raw}), locally sparse estimator for model (\ref{model}) with asynchronous observations can be obtained. Note that though we consider generalized varying coefficient model with one covariate here, it can be easily extended to the cases with more covariates.

\subsection{Sparseness penalty}\label{SecSpar}

In this section, we introduce the sparseness penalty that is utilized in (\ref{EE_raw}). We generalized the functional SCAD penalty in \citep{lin2017locally} to achieve local sparsity of $\beta_1(t)$. To be specific, sparseness penalty imposed on $\beta_1(t)$ is defined as
\begin{align}
\mathcal{L}(\beta_1) = \frac{K + 1}{2T} \int_{\mathcal{T}} p_{\lambda}(|\beta_1(t)|) dt \approx \frac{1}{2} \sum_{m = 1}^{K + 1} p_{\lambda} \Bigg ( \sqrt{ \frac{K + 1}{T} \int_{\tau_{m - 1}}^{\tau_m} \beta_1^2(t) dt } \Bigg ), \label{fSCAD_beta}
\end{align}
where $T$ is the length of $\mathcal{T}$, $\tau_m$ is the knot of the used B-spline basis, and $p_{\lambda}(\cdot)$ is the SCAD function suggested in \citep{fan2001variable}. We then transform (\ref{fSCAD_beta}) to the penalty of $\bm{\gamma}$ for the sake of computation. Let $\|\beta_{1[m]} \|_2^2 = \int_{\tau_{m - 1}}^{\tau_m} \beta_1^2(t) dt $. By local quadratic approximation $p_{\lambda}(|v|) \approx p_{\lambda}(|v_0|) + \frac{1}{2} \{p_{\lambda}^{'}(|v_0|)/|v_0|\}(v^2 - v_0^2)$ in \citep{fan2001variable}, we have
\begin{align}
&\sum_{m = 1}^{K + 1} p_{\lambda} \Bigg ( \sqrt{\frac{K + 1}{T}} \|\beta_{1[m]} \|_2 \Bigg ) \nonumber \\
\approx& \sum_{m = 1}^{K + 1} \Bigg \{ p_{\lambda} \Bigg ( \sqrt{\frac{K + 1}{T}} \|\beta_{1[m]}^{(0)} \|_2 \Bigg ) + \frac{1}{2} \frac{p_{\lambda}^\prime \Big ( \sqrt{\frac{K + 1}{T}} \|\beta_{1[m]}^{(0)} \|_2 \Big )}{\sqrt{\frac{K + 1}{T}} \|\beta_{1[m]}^{(0)} \|_2} \Bigg ( \frac{K + 1}{T} \|\beta_{1[m]} \|_2^2 - \frac{K + 1}{T} \|\beta_{1[m]}^{(0)} \|_2^2 \Bigg ) \Bigg \} \nonumber \\
=& \frac{1}{2} \sum_{m = 1}^{K + 1} \sqrt{\frac{K + 1}{T}} p_{\lambda}^\prime \Bigg ( \sqrt{\frac{K + 1}{T}} \|\beta_{1[m]}^{(0)} \|_2 \Bigg ) \frac{\|\beta_{1[m]}\|^2}{\|\beta_{1[m]}^{(0)}\|^2} + C = \sum_{m = 1}^{K + 1} \bm{\gamma}^{(0) \top} \textbf{U}_m \bm{\gamma}^{(0)} + C \nonumber \\
=& \bm{\gamma}^\top \textbf{U} \bm{\gamma} + C, \nonumber
\end{align}
where
\begin{align}
\textbf{U}_m &= \sqrt{\frac{K + 1}{T}} \frac{p_{\lambda}^\prime \Big ( \sqrt{\frac{K + 1}{T}} \|\beta_{1[m]}^{(0)} \|_2 \Big )}{2\|\beta_{1[m]}^{(0)} \|_2} \textbf{T}_m, \nonumber \\
\textbf{T}_m &= \int_{\tau_{m -1}}^{\tau_{m}} \textbf{B}(t) \textbf{B}(t)^\top dt, \quad \textbf{U} = \mbox{diag} \Big ( \textbf{O}, \sum_{m = 1}^{K + 1} \textbf{U}_m \Big ), \label{Comp_U} \\
C &= \sum_{m = 1}^{K + 1} p_{\lambda} \Bigg ( \sqrt{\frac{K + 1}{T}} \|\beta_{1[m]}^{(0)} \|_2 \Bigg ) - \frac{1}{2} \sum_{m = 1}^{K + 1} \sqrt{\frac{K + 1}{T}} p_{\lambda}^\prime \Bigg ( \sqrt{\frac{K + 1}{T}} \|\beta_{1[m]}^{(0)} \|_2 \Bigg ) \|\beta_{1[m]}^{(0)} \|_2, \nonumber
\end{align}
and $\textbf{O}$ is a $L \times L$ matrix with all elements being zero.
Here $\|\beta_{1[m]}^{(0)} \|_2$ is obtained from the initial value or the estimate in the previous iteration. Then the sparseness penalty in (\ref{EE_raw}) can be expressed as
\begin{align}
\mbox{PEN}_{\lambda}(\bm{\gamma}) = \frac{1}{2} \bm{\gamma}^\top \textbf{U} \bm{\gamma}. \nonumber
\end{align}
Here the value of $\textbf{U}$ depends on the value of $\|\beta_{1[m]}^{(0)} \|_2$, so it will be varied in the iteration process that is introduced in Section \ref{SecAlgorithm}.

\subsection{Algorithm}\label{SecAlgorithm}

We generalize the IRLS algorithm to solve our estimating equation proposed in Section \ref{SecEE}. To this end, we first rewrite (\ref{EE_raw}) into matrix form and more notations need to be introduced. Let $\wt{\textbf{X}}_i^{\star} = (\wt{\textbf{X}}_i^{\star}(S_{i1}), \ldots, \wt{\textbf{X}}_i^{\star}(S_{iM_i}))^\top$, $\wt{\textbf{X}}^{\star} = (\textbf{1}_{L_1}^\top \otimes \wt{\textbf{X}}_1^{\star \top}, \ldots, \textbf{1}_{L_n}^\top \otimes \wt{\textbf{X}}_n^{\star \top})^\top$, $\textbf{Y}_i = (Y_i(T_{i1}), \ldots, Y_i(T_{iL_i}))^\top$, $\textbf{Y} = (\textbf{Y}_1^\top \otimes \textbf{1}_{M_1}^\top, \ldots, \textbf{Y}_n^\top \otimes \textbf{1}_{M_n}^\top)^\top$, $\bm{\eta} = \wt{\textbf{X}}^{\star} \bm{\gamma}$, $\textbf{Z} = \bm{\eta} + \{\textbf{Y} - g(\bm{\eta})\} \cdot f^{\prime}\{g(\bm{\eta})\}$, $\textbf{W} = \mbox{diag}\{K_h(T_{11} - S_{11}), \ldots, K_h(T_{11} - S_{1M_1}), K_h(T_{12} - S_{11}), \ldots, K_h(T_{nL_n} - S_{nM_n})\}$ and $\textbf{H} = \mbox{diag}[1/f^{\prime}\{g(\bm{\eta})\}]$, where $\otimes$ is the Kronecker product, $\textbf{1}_{L_i}$ and $\textbf{1}_{M_i}$ are the vectors of length $L_i$ and $M_i$ with all elements being $1$, and $f^{\prime}(\cdot)$ is the first derivative of $f(\cdot)$ which is the inverse function of $g(\cdot)$. Then the penalized kernel-weighted estimating equation (\ref{EE_raw}) becomes
\begin{align}
U_n(\bm{\gamma}) = \frac{1}{N_0}\wt{\textbf{X}}^{\star \top} \textbf{W} \textbf{H} (\textbf{Z} - \bm{\eta}) - \textbf{V}_{\rho_0, \rho_1} \bm{\gamma} - \textbf{U} \bm{\gamma} = \textbf{0}, \label{EE_matrix}
\end{align}
where $\textbf{H}$, $\textbf{Z}$, $\bm{\eta}$ and $\textbf{U}$ are computed by initial value of $\bm{\gamma}$ or its estimate in the previous iteration. Through (\ref{EE_matrix}), the new estimate can be obtained by
\begin{align}
\wh{\bm{\gamma}} = (\wt{\textbf{X}}^{\star \top} \textbf{WH} \wt{\textbf{X}}^{\star} + N_0 \textbf{V}_{\rho_1, \rho_2} + N_0 \textbf{U})^{-1} \wt{\textbf{X}}^{\star \top} \textbf{WH} \textbf{Z}. \label{gamma_est}
\end{align}
Moreover, refer to \citep{lin2017locally} and \citep{zhong2021sparse}, the small elements of $\wh{\bm{\gamma}}$ are shrunk to zero in the iteration so that $\wt{\textbf{X}}^{\star \top} \textbf{WH} \wt{\textbf{X}}^{\star} + N_0 \textbf{V}_{\rho_1, \rho_2} + N_0 \textbf{U}$ would not be singular. Then the estimates of $\beta_0(t)$ and $\beta_1(t)$ are given by
\begin{align}
\wh{\beta}_0(t) = \textbf{B}(t)^\top \wh{\bm{\gamma}}^{{(0)}}, \quad \wh{\beta}_1(t) = \textbf{B}(t)^\top \wh{\bm{\gamma}}^{{(1)}}, \label{beta_est}
\end{align}
where $\wh{\bm{\gamma}}^{{(0)}}$ and $\wh{\bm{\gamma}}^{{(1)}}$ are obtained from the final estimate of $\bm{\gamma}$ via the definition $\bm{\gamma} = (\bm{\gamma}^{{(0)}\top}, \bm{\gamma}^{{(1)}\top})^\top$.

Further, the whole algorithm is summarized as follows:
\begin{itemize}[leftmargin = 35pt]

\item[Step 1:] Give initial value of $\bm{\gamma}$. Let $\bm{\gamma}^{[0]}$ denote the initial value. Here a least squares estimate with kernel weight is used, and the roughness penalty is also considered in the initial estimate, that is $\bm{\gamma}^{[0]} = (\wt{\textbf{X}}^{\star \top} \textbf{W} \wt{\textbf{X}}^{\star} + N_0 \textbf{V}_{\rho_1, \rho_2} )^{-1} \wt{\textbf{X}}^{\star \top} \textbf{W} \textbf{Y}$.

\item[Step 2:] Start with $q = 1$, for the $q$-th iteration,
    \begin{itemize}

    \item[(1)] $\bm{\eta}^{[q]} = \wt{\textbf{X}}^{\star} \bm{\gamma}^{[q - 1]}$.
    \item[(2)] $\textbf{Z}^{[q]} = \bm{\eta}^{[q]} + \{\textbf{Y} - g(\bm{\eta}^{[q]})\} \cdot f^{\prime}\{g(\bm{\eta}^{[q]})\}$ and $\textbf{H}^{[q]} = \mbox{diag} [1/f^{\prime}\{g(\bm{\eta}^{[q]})\}]$.
    \item[(3)] Compute $\textbf{U}^{[q]}$ by (\ref{Comp_U}).
    \item[(4)] $\bm{\gamma}^{[q]} = (\wt{\textbf{X}}^{\star \top} \textbf{W} \textbf{H}^{[q]} \wt{\textbf{X}}^{\star} + N_0 \textbf{V}_{\rho_1, \rho_2} + N_0 \textbf{U}^{[q]})^{-1} \wt{\textbf{X}}^{\star \top} \textbf{W} \textbf{H}^{[q]} \textbf{Z}^{[q]}$ according to (\ref{gamma_est}).
    \item[(5)] Repeat Step 2(1)-(4) until convergence.

    \end{itemize}

\item[Step 3:] Let $\wh{\bm{\gamma}} = \bm{\gamma}^{[q]}$, then compute $\wh{\beta}_0(t)$ and $\wh{\beta}_1(t)$ by (\ref{beta_est}).

\end{itemize}

\subsection{Selection of tuning parameters}\label{SecTuning}

In this section, we discuss the selection of tuning parameters involved in the computation, including the roughness parameters, sparseness parameter and the number of B-spline basis. For clarity, let $\rho_0 = \rho_1 \triangleq \wt{\rho}$, which means $\beta_0(t)$ and $\beta_1(t)$ share the same roughness parameter. However, our selecting criterion can be easily extended to the case where $\rho_0 \neq \rho_1$.

The roughness parameter $\wt{\rho}$ and the sparseness parameter $\lambda$ are jointly considered. We generalize $\mbox{EBIC}$ in \citep{chen2008extended, chen2012extended} to make it adapt to the asynchronous observations. More specifically, define
\begin{align}
\mbox{EBIC} (\wt{\rho}, \lambda) = \log (\mbox{Dev}) + df \cdot \log(n_0)/n_0 + \nu \cdot df \cdot \log (2L) /n_0, \label{EBIC}
\end{align}
where $\mbox{Dev}$ represents deviance of the estimate, $df$ is the degree of freedom, $n_0 = \#\{K_h(T_{ij} - S_{ik}) \neq 0, i = 1, \ldots, n; j = 1, \ldots L_i; k = 1, \ldots, M_i\}$ and $0 \leq \nu \leq 1$. We use $\nu = 0.5$ as suggested by \citet{huang2010variable}. Moreover, $\mbox{Dev}$ is given by
\begin{align}
\mbox{Dev} = -2 \sum_{i = 1}^n \sum_{j = 1}^{L_i} \sum_{k = 1}^{M_i} \{ Y_i(T_{ij}) \wh{\theta}_{ik} - b(\wh{\theta}_{ik}) \} K_h(T_{ij} - S_{ik}), \nonumber
\end{align}
where $\wh{\theta}_{ik} = g(\wh{Y}_i (S_{ik}))$. Then with the neglect of some constant, we have for Gaussian response,
\begin{align}
\mbox{Dev} = \sum_{i = 1}^n \sum_{j = 1}^{L_i} \sum_{k = 1}^{M_i} \{Y_i(T_{ij}) - \wh{Y}_i(S_{ik}) \}^2 K_h(T_{ij} - S_{ik}), \nonumber
\end{align}
while for Bernoulli response,
\begin{align}
\mbox{Dev} = 2 \sum_{i = 1}^n \sum_{j = 1}^{L_i} \sum_{k = 1}^{M_i} \Big [ Y_i(T_{ij}) \log \frac{Y_i (T_{ij})}{\wh{Y}_i (S_{ik})} + \{1 - Y_i(T_{ij}) \} \log \frac{1 - Y_i(T_{ij})}{1 - \wh{Y}_i(S_{ik})} \Big ] K_h(T_{ij} - S_{ik}), \nonumber
\end{align}
and for Poisson response,
\begin{align}
\mbox{Dev} = 2 \sum_{i = 1}^n \sum_{j = 1}^{L_i} \sum_{k = 1}^{M_i} [\wh{Y}_i(S_{ik}) - Y_i(T_{ij}) \log \{\wh{Y}_i(S_{ik})\}] K_h(T_{ij} - S_{ik}). \nonumber
\end{align}
Furthermore, $df$ is computed by
\begin{align}
df = \mbox{tr} \{ \wt{\textbf{X}}^{\star}_{\mathcal{A}} (\wt{\textbf{X}}^{\star \top}_{\mathcal{A}} \textbf{W}_{\mathcal{A}} \wt{\textbf{X}}^{\star}_{\mathcal{A}} + N_0 \textbf{V}_{\rho_1, \rho_2 \mathcal{A}} )^{-1} \wt{\textbf{X}}^{\star \top}_{\mathcal{A}} \textbf{W}_{\mathcal{A}} \}, \nonumber
\end{align}
where $\mathcal{A}$ is a set indexing the nonzero elements in $\wh{\bm{\gamma}}$. For the third term in (\ref{EBIC}), $2L$ is the length of $\bm{\gamma}$, and if more covariates are considered, it should be varied accordingly.

We choose the number of B-spline basis functions through CV. For a given $L$, we first select the best $\wt{\rho}$ and $\lambda$ via $\mbox{EBIC}$, and then the CV score is calculated by the same computing method as $\mbox{Dev}$ when facing response with various distributions. The effect of $L$ is discussed in our simulation study in Section \ref{SecSimL}.

\section{Theoretical results}\label{SecTheory}

We study the asymptotic properties, including consistency and sparsistency, of our estimator in this section. Let $\eta(t, \bm{\beta}) = \beta_0(t) + X(t) \beta_1(t)$, where $\bm{\beta}(t) = (\beta_0(t), \beta_1(t))^\top$. Let $\bm{\beta}_0 (t)$ be the true value of $\bm{\beta} (t)$. Define $\textbf{X}^{\star}(t) = (1, X(t))^\top$. Let $\mbox{var} \{Y(t) | X(t)\} = \sigma\{t, X(t)\}^2$ and $\mbox{cov}\{Y(s), Y(t) | X(s), X(t)\} = r\{s, t, X(s), X(t)\}$. Moreover, denote $\mbox{NULL}(f) = \{t \in \mathcal{T} : f(t) = 0\}$ and $\mbox{SUPP}(f) = \{t \in \mathcal{T} : f(t) \neq 0\}$ for any function $f(t)$. The needed assumptions are listed as follows:
\begin{assumption}\label{AssBspline}
There exists some constant $c > 0$ such that $|\beta_0^{(p^{\prime})} (t_1) - \beta_0^{(p^{\prime})} (t_2)| \leq c |t_1 - t_2|^{\nu}$ and $|\beta_1^{(p^{\prime})} (t_1) - \beta_1^{(p^{\prime})} (t_2)| \leq c |t_1 - t_2|^{\nu}, \nu \in [0, 1]$. Let $r = p^{\prime} + \nu$ and assume that $3/2 < r \leq d$, where $d$ is the degree of the B-spline basis.
\end{assumption}
\begin{assumption}\label{AssCount}
The counting process $N_i(t, s)$ is independent of $(Y_i, X_i)$ and $E\{dN_i(t, s)\} = \lambda(t, s) dt ds$, where $\lambda(t, s)$ is a bounded twice-continuous differentiable function for any $t, s \in \mathcal{T}$. Borel measure for $\mathcal{G} = \{ \lambda(t, t) > 0, t \in \mathcal{T} \}$ is strictly positive. Moreover, $P\{dN(t_1, t_2) = 1 | N(s_1, s_2) - N(s_1-, s_2-) = 1 \} = f(t_1, t_2, s_1, s_2) dt_1 dt_2$ for $t_1 \neq s_1$ and $t_2 \neq s_2$, where $f(t_1, t_2, s_1, s_2)$ is continuous and $f(t_1 \pm, t_2 \pm, s_1 \pm, s_2 \pm)$ exists.
\end{assumption}
\begin{assumption}\label{AssSparse}
The tuning parameter $\lambda \rightarrow 0$ as $n \rightarrow \infty$. Assume that $\sqrt{\int_{\mbox{SUPP}(\beta_1)} p_{\lambda}^{\prime} (|\beta_1 (t)|)^2 dt} = O(n^{-1/2} K^{-3/2})$, $\sqrt{\int_{\mbox{SUPP}(\beta_1)} p_{\lambda}^{\prime \prime} (|\beta_1(t)|)^2 dt} = o(K^{-3/2})$.
\end{assumption}
\begin{assumption}\label{AssDiff}
For any $\bm{\beta}$ in a neighborhood of $\bm{\beta}_0$, we assume that $E [ \textbf{X}^{\star}(s) g \{ \eta(t, \bm{\beta}) \} ]$ and $E [ \textbf{X}^{\star}(s) g^{\prime} \{ \eta(t, \bm{\beta}) \} X^b(t) ]$ are twice-continuous differentiable for any $(t, s) \in \mathcal{T}^2$, where $b = 0, 1$. Moreover, $E[\textbf{X}^{\star}(s_1) \textbf{X}^{\star}(s_2)^\top g \{ \eta(t_1, \bm{\beta}) \} g \{ \eta(t_2, \bm{\beta}) \}]$ and $E [\textbf{X}^{\star}(s_1) \textbf{X}^{\star}(s_2)^\top r\{ t_1, t_2, X(t_1), X(t_2) \}]$ are twice-continuous differentiable for any $(t_1, t_2, s_1, s_2) \in \mathcal{T}^4$.
\end{assumption}
\begin{assumption}\label{AssBound}
For any $\bm{\beta}$ in a neighborhood of $\bm{\beta}_0$, we assume that $E[ \textbf{X}^{\star}_2(s) \textbf{X}^{\star}_2(s)^\top g^{\prime} \{ \eta(s, \bm{\beta}) \}^2 ]$ and $E[ \textbf{X}^{\star}(s) \sigma\{s, X(s)\}^2 ]$ are uniformly bounded in $s$, where $\textbf{X}^{\star}_2(s) = (1, X^2(t))^\top$.
\end{assumption}
\begin{assumption}\label{AssZero}
If $\psi_0$ and $\psi_1$ satisfy $\psi_0(s) + \psi_1(s) X(s) = 0, \forall s \in \mathcal{G}$ with probability $1$, then $\psi_0 = 0$ and $\psi_1 = 0$.
\end{assumption}
\begin{assumption}\label{AssTuning}
The bandwidth $h = O(n^{-1/5})$ and the number of knots $K = O(n^{\frac{4}{5(1 + 2r)}})$. Denote $\rho = \max(\rho_0, \rho_1)$, and $\rho = o(n^{-1/2})$. The sparseness parameter $\lambda = o(1)$ and $\lambda n^{1/2} K^{-1/2} h^{1/2} \rightarrow \infty$.
\end{assumption}
\begin{assumption}\label{AssKernel}
The kernel function $K(\cdot)$ is a symmetric density function. Assume that $\int z^2 K(z) dz < \infty$ and $\int K(z)^2 dz < \infty$.
\end{assumption}

Assumption \ref{AssBspline} is similar to (C2) in \citep{lin2017locally}, and this assumption is used to justify the B-spline approximation. Requirement for the counting process is presented in Assumption \ref{AssCount} and is the same as Condition 1 in \citep{cao2015regression} and Assumption 3 in \citep{li2020regression}. Assumption \ref{AssSparse} is analogous to (C3) in \citep{lin2017locally}, while Assumptions \ref{AssDiff}-\ref{AssZero} are parallel to assumptions in \citep{li2020regression}. Furthermore, Assumption \ref{AssTuning} gives the choosing condition of tuning parameters and Assumption \ref{AssKernel} is a common assumption for kernel function.

\begin{theorem}\label{TheConsis}
Under Assumptions \ref{AssBspline} - \ref{AssKernel}, we have $\sup_{t \in \mathcal{T}} |\wh{\beta}_0(t) - \beta_0(t)| = O_p(n^{-1/2}K^{1/2}h^{-1/2})$ and $\sup_{t \in \mathcal{T}} |\wh{\beta}_1(t) - \beta_1(t)| = O_p(n^{-1/2}K^{1/2}h^{-1/2})$.
\end{theorem}

The above theorem states the consistency of both $\beta_0(t)$ and $\beta_1(t)$, and the convergence rates are also given. Based on that, we discuss the sparsistency of $\beta_1(t)$ in the following theorem.

\begin{theorem}\label{TheSpar}
Under Assumptions \ref{AssBspline} - \ref{AssKernel}, $\mbox{NULL}(\wh{\beta}_1) \rightarrow \mbox{NULL}(\beta_1)$ and $\mbox{SUPP}(\wh{\beta}_1) \rightarrow \mbox{SUPP}(\beta_1)$ in probability, as $n \rightarrow \infty$.
\end{theorem}

According to Theorem \ref{TheSpar}, the zero-valued subintervals of our estimate $\wh{\beta}_1(t)$ are consistent with the true zero-valued subintervals. The proofs of Theorem \ref{TheConsis} and Theorem \ref{TheSpar} are relegated to the Supplementary Material.

\section{Simulation studies}\label{SecSim}

\subsection{Numerical performance}\label{SecNum}

In this section, we discuss the performance of the proposed method through simulation studies. The simulated datasets are generated from model (\ref{model}), and Gaussian response, Bernoulli response and Poisson response are all in consideration. Moreover, for each distribution, both nonsparse coefficient function and coefficient function with local sparsity are taken into account. The detailed settings are as follows:
\begin{itemize}

\item Gaussian cases: The intercept function is set as $\beta_0(t) = \cos(2 \pi t), t \in [0, 1]$. For the nonsparse setting, the coefficient function $\beta_1(t) = \sin (2 \pi t)$, while for the sparse setting, $\beta_1(t) = 2 \cdot \{ B_6(t) + B_7(t) \}$, where $B_l (t)$ is the $l$-th B-spline basis on $[0, 1]$ with degree three and nine equally spaced interior knots. We generate the covariate functions in the same way as \citet{lin2017locally}, that is $X_{i} (t) = \sum_{l = 1} a_{il} B_l^X(t)$, where $a_{ij}$ is obtained from the standard normal distribution and $B_l^X(t)$ is the $l$-th B-spline basis on $[0, 1]$ with degree four and $69$ equally spaced interior knots. The sample size is set as $n = 200$. Then $Y_i(t)$ is generated from Gaussian distribution with mean $\beta_0(t) + \beta_1(t) X_i(t)$ and standard error one. To get asynchronous data, the observation sizes of response and covariate are generated independently from Poisson distribution with one additional observation to avoid the cases with no measurement. Here response and covariate share the same intensity rate $m$, and $m$ is set to be $15$ and $20$. Then the observation times are uniformly selected on $[0, 1]$.

\item Bernoulli cases: The settings are the same as Gaussian cases, except that $Y_i(t)$ is generated from Bernoulli distribution with mean $\beta_0(t) + \beta_1(t) X_i(t)$.

\item Poisson cases: The settings are the same as Gaussian cases, except that $Y_i(t)$ is generated from Poisson distribution with mean $\beta_0(t) + \beta_1(t) X_i(t)$.

\end{itemize}

The proposed LocKer method is compared with other four approaches in the simulation. The first one is a reconstruction method, that is synchronizing the response and covariate by PACE \citep{yao2005functional} like \citet{csenturk2013modeling}, and then employing the traditional IRLS algorithm. The moment method in \citep{csenturk2013modeling}, the approach in \citet{cao2015regression} and the penalized least squares estimating (PLSE) method investigated by \citet{tu2020estimation} are also considered. However, \citet{tu2020estimation} investigated local sparse estimator for varying coefficient model with synchronous observation. So to implement their method for asynchronous cases, we first synchronize the data by smoothing and then apply PLSE to the synchronized data. These four methods are denoted as Recon, Moment, Cao, PLSE respectively for simplicity. Though Cao method is available for regression model with Bernoulli and Poisson response, it is quite slow for these non-Gaussian cases since it is a pointwise method. Hence, identity link is used for Cao method in all considered cases. Moreover, PLSE is only applicable to regression model with Gaussian response, so the responses are seen as to be Gaussian distributed for PLSE in all cases.

We evaluate the integrated square error (ISE) of the estimated intercept function and coefficient function for each method. To be specific,
\begin{align}
&\mbox{ISE}_0 = \int_{\mathcal{T}} \{\wh{\beta}_0(t) - \beta_0(t) \}^2 dt, \nonumber \\
&\mbox{ISE}_1 = \int_{\mathcal{T}} \{\wh{\beta}_1(t) - \beta_1(t) \}^2 dt. \nonumber
\end{align}
In the simulation, $100$ runs are conducted for each cases. The average $\mbox{ISE}$ and the standard error are compared among various methods.

Table \ref{SimGaussian} reports the averaged $\mbox{ISE}_0$ and $\mbox{ISE}_1$ of Gaussian cases. With various settings for coefficient function $\beta_1(t)$ and observation rate $m$, the simulation results show similar trend. For the estimation of intercept function $\beta_0(t)$, all these five methods give promising results with minor difference on $\mbox{ISE}_0$. On the other hand, it is evident that our LocKer method exhibits significant advantages for the estimation of $\beta_1(t)$, regardless the true $\beta_1(t)$ is sparse or not. These results demonstrate that synchronizing approach and pointwise approach are not adequate enough, which further indicates the importance of using observed data directly and taking sufficient account of smoothness in estimation. Moreover, it can be seen that more precise estimating results are obtained for each method with the increase of observation rate.

Simulation results for Bernoulli cases are presented in Table \ref{SimBern}. The $\mbox{ISE}_0$ and $\mbox{ISE}_1$ are observed to be higher compared with the errors in Gaussian cases, which implies that Bernoulli response is more difficult to handle. However, the proposed LocKer still outperforms the other four methods in estimating $\beta_1(t)$ for both nonsparse and sparse settings, though Recon and Moment methods are slightly better in estimating $\beta_0(t)$. The reason for the invalid behaviour of Cao and PLSE methods is that they simply treat the Bernoulli response as Gaussian response here.
Furthermore, Table \ref{SimPoi} displays the simulation results for Poisson cases. We can find that the proposed LocKer achieves the most accurate estimates for both $\beta_0(t)$ and $\beta_1(t)$ in each considered setting.

In summary, our LocKer method yields encouraging estimating results for each cases in comparison with all the other methods. We conjecture the superiority of our method is due to the employment of FDA approach and kernel technique, as well as the consideration of local sparsity.

\begin{table}[H]
\caption{The averaged $\mbox{ISE}_0$ and $\mbox{ISE}_1$ across $100$ runs for five methods in Gaussian cases, with standard deviation in parentheses.}
\label{SimGaussian}
\begin{center}
\setlength{\tabcolsep}{1mm}{
\begin{tabular}{cccccc}
\hline
 & & \multicolumn{2}{c}{$n = 200, m = 15$} & \multicolumn{2}{c}{$n = 200, m = 20$} \\
 & & $\mbox{ISE}_0$ & $\mbox{ISE}_1$& $\mbox{ISE}_0$ & $\mbox{ISE}_1$ \\
\hline
\multirow{5}{*}{Nonsparse}&Recon&0.0050 (0.0022)&0.2768 (0.0505)&0.0044 (0.0019)&0.1889 (0.0455)\\
 &Moment&0.0045 (0.0022)&0.4154 (0.1826)&0.0033 (0.0017)&0.4001 (0.0581)\\
 &Cao&0.0072 (0.0031)&0.3000 (0.0326)&0.0059 (0.0028)&0.2841 (0.0344)\\
 &PLSE&0.0244 (0.0106)&0.3994 (0.0839)&0.0145 (0.0066)&0.2966 (0.0998)\\
 &LocKer&0.0170 (0.0081)&0.0385 (0.0255)&0.0094 (0.0062)&0.0217 (0.0148)\\
\hline
\multirow{5}{*}{Sparse}&Recon&0.0049 (0.0025)&0.2329 (0.0713)&0.0045 (0.0023)&0.1578 (0.0516)\\
 &Moment&0.0052 (0.0059)&0.5350 (0.2588)&0.0033 (0.0016)&0.4972 (0.0648)\\
 &Cao&0.0071 (0.0035)&0.3176 (0.0627)&0.0057 (0.0033)&0.3124 (0.0514)\\
 &PLSE&0.0216 (0.0081)&0.3025 (0.0992)&0.0153 (0.0057)&0.2147 (0.0780)\\
 &LocKer&0.0131 (0.0075)&0.0515 (0.0303)&0.0087 (0.0043)&0.0302 (0.0173)\\
\hline
\end{tabular}}
\end{center}
\end{table}

\begin{table}[H]
\caption{The averaged $\mbox{ISE}_0$ and $\mbox{ISE}_1$ across $100$ runs for five methods in Bernoulli cases, with standard deviation in parentheses.}
\label{SimBern}
\begin{center}
\setlength{\tabcolsep}{1mm}{
\begin{tabular}{cccccc}
\hline
 & & \multicolumn{2}{c}{$n = 200, m = 15$} & \multicolumn{2}{c}{$n = 200, m = 20$} \\
 & & $\mbox{ISE}_0$ & $\mbox{ISE}_1$& $\mbox{ISE}_0$ & $\mbox{ISE}_1$ \\
\hline
\multirow{5}{*}{Nonsparse}&Recon&0.0128 (0.0061)&0.3123 (0.0824)&0.0106 (0.0057)&0.2264 (0.0791)\\
 &Moment&0.0171 (0.0085)&0.6108 (0.3848)&0.0131 (0.0064)&0.4744 (0.2760)\\
 &Cao&0.5600 (0.0139)&0.4530 (0.0133)&0.5590 (0.0135)&0.4480 (0.0142)\\
 &PLSE&0.5132 (0.0170)&0.4856 (0.0195)&0.5163 (0.0150)&0.4721 (0.0255)\\
 &LocKer&0.0531 (0.0267)&0.1777 (0.0973)&0.0332 (0.0155)&0.1074 (0.0578)\\
\hline
\multirow{5}{*}{Sparse}&Recon&0.0182 (0.0075)&0.2898 (0.0966)&0.0172 (0.0067)&0.2444 (0.0892)\\
 &Moment&0.0230 (0.0113)&0.6906 (0.3372)&0.0193 (0.0074)&0.5646 (0.1204)\\
 &Cao&0.5751 (0.0150)&0.5259 (0.0148)&0.5753 (0.0113)&0.5239 (0.0140)\\
 &PLSE&0.5272 (0.0175)&0.5490 (0.0301)&0.5311 (0.0119)&0.5381 (0.0331)\\
 &LocKer&0.0426 (0.0235)&0.2600 (0.1094)&0.0291 (0.0147)&0.1773 (0.0805)\\
\hline
\end{tabular}}
\end{center}
\end{table}

\begin{table}[H]
\caption{The averaged $\mbox{ISE}_0$ and $\mbox{ISE}_1$ across $100$ runs for five methods in Poisson cases, with standard deviation in parentheses.}
\label{SimPoi}
\begin{center}
\setlength{\tabcolsep}{1mm}{
\begin{tabular}{cccccc}
\hline
 & & \multicolumn{2}{c}{$n = 200, m = 15$} & \multicolumn{2}{c}{$n = 200, m = 20$} \\
 & & $\mbox{ISE}_0$ & $\mbox{ISE}_1$& $\mbox{ISE}_0$ & $\mbox{ISE}_1$ \\
\hline
\multirow{5}{*}{Nonsparse}&Recon&0.0257 (0.0078)&0.2789 (0.0573)&0.0234 (0.0064)&0.1929 (0.0437)\\
 &Moment&0.0285 (0.0083)&0.3335 (0.1157)&0.0253 (0.0067)&0.3597 (0.0489)\\
 &Cao&1.9949 (0.1056)&0.2645 (0.0378)&1.9772 (0.0794)&0.2496 (0.0371)\\
 &PLSE&1.6426 (0.1044)&0.3555 (0.0909)&1.7170 (0.0942)&0.2408 (0.0773)\\
 &LocKer&0.0163 (0.0103)&0.0345 (0.0186)&0.0096 (0.0069)&0.0192 (0.0128)\\
\hline
\multirow{5}{*}{Sparse}&Recon&0.0660 (0.0166)&0.2462 (0.0940)&0.0660 (0.0146)&0.1670 (0.0903)\\
 &Moment&0.0730 (0.0234)&0.4579 (0.0962)&0.0745 (0.0175)&0.4791 (0.0647)\\
 &Cao&1.8242 (0.0954)&0.4116 (0.0511)&1.8220 (0.0752)&0.3991 (0.0496)\\
 &PLSE&1.4866 (0.0916)&0.4303 (0.1172)&1.5611 (0.0819)&0.3346 (0.1184)\\
 &LocKer&0.0268 (0.0128)&0.0912 (0.0604)&0.0185 (0.0097)&0.0465 (0.0225)\\
\hline
\end{tabular}}
\end{center}
\end{table}

\subsection{The effect of $L$}\label{SecSimL}

We mainly discuss the accuracy of estimation in Section \ref{SecNum}. In this section, we tend to explore how the number of B-spline basis functions influences the estimation, especially the ability of identifying zero-valued subintervals of $\beta_1(t)$. Since local sparsity is also taken into account for PLSE method, we also consider the comparison with PLSE in this section. The setups are the same as the settings in Section \ref{SecNum}, except that the response and covariate are set to be observed at the same times to make the comparison with PLSE more meaningful. To quantify the identifying ability, we compute the values of $\beta_1(t)$ and $\wh{\beta}_1(t)$ at a sequence of dense grids on $[0, 1]$, and calculate the rates of grids that correctly identified to be zero and falsely estimated to be zero, which are denoted by $\mbox{TP}$ and $\mbox{FN}$ respectively. Moreover, the closer $\mbox{TP}$ is to $1$ and the closer $\mbox{FN}$ is to $0$, the better the identifying ability is.

Tables \ref{SimGaussianm15}-\ref{SimGaussianm20} list the simulation results with the use of different values of $L$ in Gaussian cases. For the nonsparse settings, $\mbox{ISE}_0$ and $\mbox{ISE}_1$ of the proposed LocKer method decrease with the increase of $L$, and they are better than that of PLSE. Moreover, $\mbox{TP}$ does not exist for nonsparse settings, so only $\mbox{FN}$ is reported. Here both methods achieve zero-valued $\mbox{FN}$, which means no grid is falsely identified, indicating that subintervals can be effectively identified for coefficient function without local sparsity by both methods.

For the sparse settings, while the estimation of $\beta_0(t)$ becomes better with the growth of $L$, we find that both methods give the best estimation of $\beta_1(t)$ when $L = 13$. The reason is related to the setting of $\beta_1(t)$. Recall that to ensure local sparsity of $\beta_1(t)$, we utilize the B-spline basis with degree three and nine equally spaced interior knots in the setup. Therefore, B-spline basis used in the setup is coincided with B-spline basis applied in the estimation, which leads to the nice performance of our method for $L = 13$. Except the cases where $L = 13$, it is obvious that larger value of $L$ can make better estimation in terms of both accuracy and identifying ability. Compared with PLSE, our method produces more precise estimate for $\beta_0(t)$, but $\mbox{ISE}_1$ is slightly higher than that of PLSE. However, as for the identifying ability, the proposed LocKer method is much better than PLSE according to both $\mbox{TP}$ and $\mbox{FN}$, which shows the advantage of our method in zero-valued subintervals identification.

To sum up, though larger value of $L$ is beneficial for the identification in some general cases, it should be realized that more B-spline basis functions would bring more parameters in the estimation, thus increase the difficulty of estimation. Furthermore, discussion about the results in Bernoulli and Poisson cases is provided in the Supplementary Material.

\begin{table}[H]
\caption{The averaged $\mbox{ISE}_0$, $\mbox{ISE}_1$, $\mbox{TP}$ and $\mbox{FN}$ across $100$ runs for PLSE and LocKer using various values of $L$ when $n = 200, m = 15$ in Gaussian cases, with standard deviation in parentheses.}
\label{SimGaussianm15}
\begin{center}
\setlength{\tabcolsep}{1mm}{
\begin{tabular}{ccccccc}
\hline
 & & & $\mbox{ISE}_0$ & $\mbox{ISE}_1$ & $\mbox{TP}$ & $\mbox{FN}$ \\
\hline
\multirow{4}{*}{$L = 10$}&\multirow{2}{*}{Nonsparse}&PLSE&0.0120 (0.0054)&0.0196 (0.0064)&--&0 (0)\\
 & &LocKer&0.0115 (0.0039)&0.0139 (0.0048)&--&0 (0)\\
&\multirow{2}{*}{Sparse}&PLSE&0.0209 (0.0079)&0.0159 (0.0062)&0.1740 (0.2254)&0 (0)\\
 & &LocKer&0.0099 (0.0036)&0.0169 (0.0060)&0.5564 (0.1486)&0 (0)\\
\hline
\multirow{4}{*}{$L = 13$}&\multirow{2}{*}{Nonsparse}&PLSE&0.0123 (0.0049)&0.0189 (0.0060)&--&0 (0)\\
 & &LocKer&0.0077 (0.0029)&0.0115 (0.0054)&--&0 (0)\\
&\multirow{2}{*}{Sparse}&PLSE&0.0209 (0.0075)&0.0070 (0.0049)&0.6109 (0.3012)&0 (0)\\
 & &LocKer&0.0065 (0.0031)&0.0056 (0.0041)&0.9777 (0.0625)&0 (0)\\
\hline
\multirow{4}{*}{$L = 15$}&\multirow{2}{*}{Nonsparse}&PLSE&0.0093 (0.0038)&0.0186 (0.0064)&--&0 (0)\\
 & &LocKer&0.0063 (0.0025)&0.0095 (0.0054)&--&0 (0)\\
&\multirow{2}{*}{Sparse}&PLSE&0.0152 (0.0059)&0.0081 (0.0039)&0.3925 (0.2461)&0.0230 (0.0365)\\
 & &LocKer&0.0053 (0.0022)&0.0161 (0.0072)&0.8619 (0.0461)&0.0195 (0.0359)\\
\hline
\multirow{4}{*}{$L = 20$}&\multirow{2}{*}{Nonsparse}&PLSE&0.0093 (0.0039)&0.0204 (0.0062)&--&0 (0)\\
 & &LocKer&0.0047 (0.0021)&0.0076 (0.0055)&--&0 (0)\\
&\multirow{2}{*}{Sparse}&PLSE&0.0179 (0.0065)&0.0098 (0.0049)&0.5022 (0.2323)&0.0786 (0.0619)\\
 & &LocKer&0.0043 (0.0018)&0.0135 (0.0092)&0.9086 (0.0631)&0.0042 (0.0167)\\
\hline
\end{tabular}}
\end{center}
\end{table}

\begin{table}[H]
\caption{The averaged $\mbox{ISE}_0$, $\mbox{ISE}_1$, $\mbox{TP}$ and $\mbox{FN}$ across $100$ runs for PLSE and LocKer using various values of $L$ when $n = 200, m = 20$ in Gaussian cases, with standard deviation in parentheses.}
\label{SimGaussianm20}
\begin{center}
\setlength{\tabcolsep}{1mm}{
\begin{tabular}{ccccccc}
\hline
 & & & $\mbox{ISE}_0$ & $\mbox{ISE}_1$ & $\mbox{TP}$ & $\mbox{FN}$ \\
\hline
\multirow{4}{*}{$L = 10$}&\multirow{2}{*}{Nonsparse}&PLSE&0.0065 (0.0031)&0.0128 (0.0049)&--&0 (0)\\
 & &LocKer&0.0071 (0.0027)&0.0089 (0.0044)&--&0 (0)\\
&\multirow{2}{*}{Sparse}&PLSE&0.0128 (0.0057)&0.0136 (0.0051)&0.1621 (0.2108)&0 (0)\\
 & &LocKer&0.0061 (0.0027)&0.0159 (0.0045)&0.5587 (0.1517)&0 (0)\\
\hline
\multirow{4}{*}{$L = 13$}&\multirow{2}{*}{Nonsparse}&PLSE&0.0066 (0.0029)&0.0131 (0.0045)&--&0 (0)\\
 & &LocKer&0.0045 (0.0020)&0.0075 (0.0046)&--&0 (0)\\
&\multirow{2}{*}{Sparse}&PLSE&0.0143 (0.0046)&0.0050 (0.0033)&0.6009 (0.2522)&0 (0)\\
 & &LocKer&0.0038 (0.0016)&0.0049 (0.0034)&0.9838 (0.0542)&0 (0)\\
\hline
\multirow{4}{*}{$L = 15$}&\multirow{2}{*}{Nonsparse}&PLSE&0.0056 (0.0023)&0.0134 (0.0044)&--&0 (0)\\
 & &LocKer&0.0041 (0.0018)&0.0075 (0.0043)&--&0 (0)\\
&\multirow{2}{*}{Sparse}&PLSE&0.0092 (0.0035)&0.0064 (0.0035)&0.3104 (0.2266)&0.0126 (0.0261)\\
 & &LocKer&0.0033 (0.0014)&0.0096 (0.0038)&0.8654 (0.0613)&0.0241 (0.0345)\\
\hline
\multirow{4}{*}{$L = 20$}&\multirow{2}{*}{Nonsparse}&PLSE&0.0065 (0.0024)&0.0150 (0.0047)&--&0 (0)\\
 & &LocKer&0.0035 (0.0018)&0.0057 (0.0039)&--&0 (0)\\
&\multirow{2}{*}{Sparse}&PLSE&0.0128 (0.0049)&0.0078 (0.0033)&0.5393 (0.1997)&0.0748 (0.0582)\\
 & &LocKer&0.0028 (0.0015)&0.0073 (0.0033)&0.9484 (0.0242)&0.0116 (0.0268)\\
\hline
\end{tabular}}
\end{center}
\end{table}

\section{Real data analysis}\label{SecReal}

In this section, we apply the proposed LocKer method to an AIDS study \citep{wohl2005cytomegalovirus}. In this study, 190 individuals were enrolled and they were followed from July 1997 to September 2002 with measurement of their HIV viral load and CD4 cell counts. As a results of missing visits and randomness of HIV infections, HIV viral load and CD4 cell counts were measured at various times for each individual. Hence, to explore the relationship between these two variables, we have to take into account the asynchronous feature of the data.

\begin{figure}
  \centering
  \includegraphics[width=\textwidth]{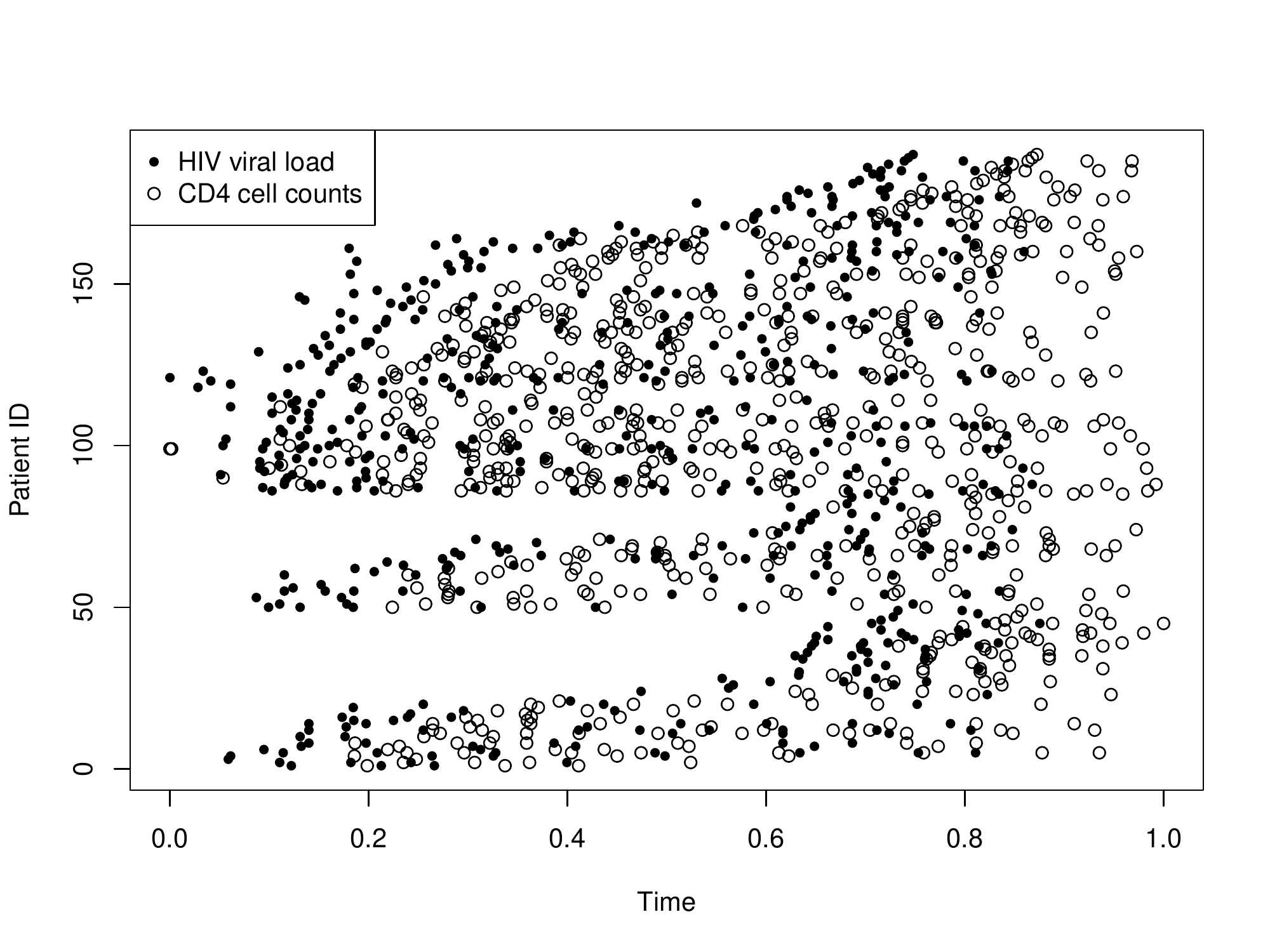}\\
  \caption{Observation times of HIV viral load and CD4 cell counts for the $181$ individuals.}
  \label{FigRealAsy}
\end{figure}

\begin{figure}[H]
  \centering
  \includegraphics[width=0.7\textwidth]{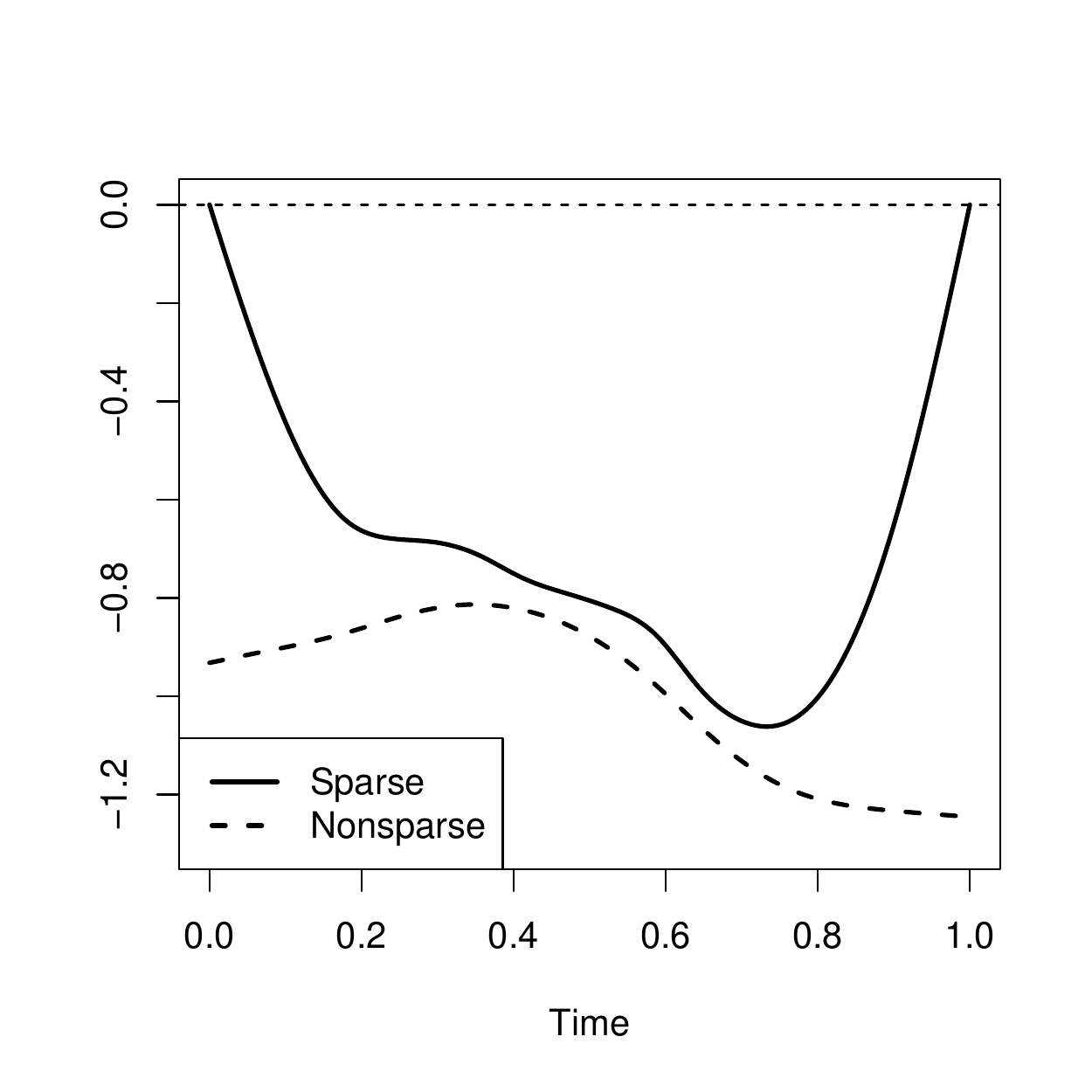}\\
  \caption{Estimates of the coefficient function in the AIDS study. Solid line is the LocKer estimate with the roughness parameter and sparseness parameter being selected by $\mbox{EBIC}$. Dashed line is the LocKer estimate with the roughness parameter being selected by $\mbox{EBIC}$ and sparseness parameter being zero.}
  \label{FigReal}
\end{figure}

Here HIV viral load is treated as response and CD4 cell counts is treated as covariate. We remove the individuals with no measurement of HIV viral load or CD4 cell counts first, and there are $181$ individuals left. As \citet{sun2021regression}, we also rescale the time interval into $\mathcal{T} = [0, 1]$. The observation times of HIV viral load and CD4 cell counts for each individual are exhibited in Figure \ref{FigRealAsy}. It is evident that the response and covariate are observed at distinct times and both variables are observed irregularly. Furthermore, Figure \ref{FigReal} shows the estimated coefficient function by LocKer in solid line, and the tuning parameters are selected as illustrated in Section \ref{SecTuning}. It can be seen that HIV viral load and CD4 cell counts are negative related, which is coincide with the discovery in medical studies. In addition, our method can also display the variation pattern of the relationship between these two variables.

To illustrate the effect of the sparseness penalty, we plot the estimated coefficient function that obtained without the use of sparseness penalty in Figure \ref{FigReal} as well. Based on the comparison of these two estimates, we find that though the sparseness penalty does not cause obvious local sparsity to the estimate, it still brings convenience to the interpretation. The LocKer estimate emphasizes the negative association at about $t = 0.8$, while the estimate obtained without sparseness penalty is much flatter. That means the imposed sparseness penalty can help to highlight the important effects, which makes the results more interpretable.

\section{Conclusion and discussion}\label{SecDiscussion}

In this paper, we employ FDA method in the estimation of generalized varying coefficient model. Moreover, kernel technique is utilized to solve the asynchronous issue and a sparseness penalty is imposed to improve accuracy and interpretability of the estimates. The theoretical study verifies both consistency and sparsistency of the proposed LocKer method. The extensive simulation experiments and practical application also suggest the encouraging performance of LocKer method.

However, the main focus of this paper is generalized varying coefficient model, which means only the response value and covariate value recorded at the same time are relevant. A more general model can be expressed as
\begin{align}
E\{Y(t)|X(s), s \in \mathcal{T} \} = g \Big \{ \beta_0(t) + \int_{\mathcal{T}} X(s) \beta_1(s, t)ds \Big \}, t \in \mathcal{T}. \nonumber
\end{align}
In the above model, the response is related to the value of covariate on the whole interval $\mathcal{T}$ rather than at one exact time point, which is more practical in the analysis of real world dataset. Furthermore, the consideration of asynchronous issue and local sparsity in this model is also quite essential but in high difficulty, which is the interest of our future research.

\section*{Supplementary Materials}

The Supplementary Material contains the proofs of Theorem \ref{TheConsis} and Theorem \ref{TheSpar}, and some additional simulation results are also provided.

\section*{Acknowledgements}

This work was supported by Public Health $\&$ Disease Control and Prevention, Major Innovation $\&$ Planning Interdisciplinary Platform for the ``Double-First Class" Initiative, Renmin University of China. This work was supported by the Outstanding Innovative Talents Cultivation Funded Programs 2022 of Renmin University of China.

\bibliographystyle{unsrtnat}
\bibliography{ref}

\end{document}